\documentclass[3p, 12pt]{elsarticle}

\usepackage{graphicx}
\usepackage{subfigure}
\usepackage{amssymb}
\usepackage[dvipdfx,cmyk]{xcolor}
\biboptions{sort&compress}

\journal{Communications in Nonlinear Science and Numerical Simulation}

\begin{document}

\title{Superstable cycles for antiferromagnetic $Q$-state Potts and three-site interaction Ising models on recursive lattices}

\author[yerphi,lavras]{N.~Ananikian}
\ead{ananik@mail.yerphi.am}

\author[artuso,artuso1]{R.~Artuso}

\author[yerphi,dijon,ashtarak]{L.~Chakhmakhchyan}
\ead{levonc@rambler.ru}

\address[yerphi]{A.I.~Alikhanyan National Science Laboratory, 0036 Yerevan, Armenia}

\address[lavras]{Departamento de Ciencias Exatas, Universidade Federal de Lavras, CP 3037, 37200-000 Lavras-MG, Brazil}

\address[artuso]{Dipartimento di Scienza e Alta Tecnologia, Universit\'{a} degli Studi dell'Insubria, Via Valleggio 11, 22100 Como, Italy}

\address[artuso1]{I.N.F.N. Sezione di Milano, Via Celoria 16, 20133 Milano. Italy}

\address[dijon]{Laboratoire Interdisciplinaire Carnot de Bourgogne, UMR CNRS 6303, \\ Universit\'{e} de Bourgogne, 21078 Dijon Cedex, France}

\address[ashtarak]{Institute for Physical Research, 0203 Ashtarak-2, Armenia}

\date{\today}
\begin{abstract}
We consider the superstable cycles of the $Q$-state Potts (QSP) and the three-site interaction antiferromagnetic Ising (TSAI) models on recursive lattices. The rational mappings describing the models' statistical properties are obtained via the recurrence relation technique. We provide analytical solutions for the superstable cycles of the second order for both models. A particular attention is devoted to the period three window. Here we present an exact result for the third order superstable orbit for the QSP and a numerical solution for the TSAI model. Additionally, we point out a non-trivial connection between bifurcations and superstability: in some regions of parameters a superstable cycle is not followed by a doubling bifurcation. { Furthermore, we use symbolic dynamics to understand the changes taking place at points of superstability and to distinguish areas between two consecutive superstable orbits}. \let\thefootnote\relax\footnote{Abbreviations: \\ QSP model: $Q$-state Potts model \\ TSAI model: three-site antiferromagnetic Ising model}
\end{abstract}


\maketitle

\section{\label{intr}Introduction}

The theory of dynamical systems plays a key role in different aspects of modern physics, ecology and economics. In particular, logistic-like maps are widely used for modeling the population evolution of living organisms with a restricted ``carrying capacity" of the environment \cite{ecol}, the evolution of corruption in public procurement \cite{corup} and for a description of economic cycles \cite{econ}. Although logistic maps belong to one of the simplest family of non-linear maps, they may exhibit quite a complicated behavior, including period doubling cascade, chaos and periodic windows \cite{schuster}.

On the other hand, the dynamical approach represents an essential tool in the theory of phase transitions and criticality \cite{baxter, gujrati, gujrati1, ananik, ananik1, ananik2} and it greatly enhanced our understanding of the phase structure and critical properties of spin and gauge models. The method is widely used to investigate exact solution of spin models on hierarchical lattices, which are good approximations for real ones (the so called Bethe-Peierls approximation) \cite{recur, recur1, recur2, jetp, andrade, andrade1}. This technique can also be applied to the generalized Bethe (Husimi) lattice, to describe properties of frustrated systems with multisite interactions, and RNA-like polymers \cite{Husimi, Husimi1, RNA}. The multisite interaction Ising and $Q$-state Potts models are of particular interest: the first one is efficient in the analysis of magnetic properties of solid $^3$He \cite{He3, He31}, while Potts model, apart of being strongly related to problems in magnetism \cite{potts, potts1, potts2}, falls in the same universality class as gelation processes in branched polymers \cite{lyubensky, anfinsen}; note that the model is well-defined for non-integer values of $Q$ (as pointed out in \cite{kast}).

In the present work we consider the three site interaction Ising and the $Q$-state Potts models ($Q<2$) on Husimi and Bethe lattices, respectively. A remarkable feature of these systems is their exact solvability through recurrence relations technique. Within this method, statistical properties of a system are associated to one- or multidimensional rational mappings \cite{recur, recur1, recur2, jetp, Husimi, Husimi1, He3, He31, gujrati, gujrati1}. In the case of antiferromagnetic coupling between lattice nodes, both models exhibit a complex behavior, featuring doubling bifurcations, chaotic regimes, intermittency,  and superstable cycles.

The aim of this paper is to study the superstability phenomenon in the models mentioned above. Our technique provides an analytical solution for superstable cycles of the second order for the Potts and Ising models. Additionally, we present an analytical solution for the superstable cycle of the third order (i.e. in the period three window) for the rational mapping associated to the $Q$-state Potts model. Note that the period three window has been mainly analyzed numerically and mostly for polynomial maps: as a matter of fact there are very few analytical solutions, and for the logistic map only (see Refs.~\cite{analytic,analytic1,analytic2,lee,lee1}): in this work we add one more exact result to this short list. Furthermore, we find a non-trivial connection of superstable cycles to bifurcations in the period three window: in some areas of parameters a superstable cycle is not followed by a doubling bifurcation (which is not the case for the logistic map).

Another goal of the work is to analyze the qualitative changes occurring at points of superstability. Previously, superstable cycles were mainly considered as an auxiliary tool for the search or analysis of stable cycles. Moreover, to the best of our knowledge, the problem of superstability was examined for quadratic maps only. In particular,
efforts have been devoted to find the third order superstable cycle of the logistic map: the problem was firstly discussed by Guckenheimer et al. in~\cite{ecol} and was solved analytically in~\cite{lee}.
%Additionally, the superstable cycles were proved to follow the Sharkovskii's order for a family of one-parameter endomorphisms of the real line \cite{shark, shark1}.
In what follows we also make use of symbolic dynamics \cite{dav} and point out how it evolves at points of superstability of the above mentioned rational mappings. We show how ``phases" between two consecutive superstable cycles can be distinguished by means of generic symbolic sequences.

The paper is organized as follows: in the next section we give a brief description of the Potts and Ising models on recursive lattices and provide their exact solution. In Section~\ref{sup} we address the superstability phenomenon and the period three window. The symbolic dynamics and its connection to the superstability is discussed in Section~\ref{symb}. Finally, in Section~\ref{concl} we give our concluding remarks.

\section{\label{model}Models and their Exact Solution}
\begin{figure}[ht]
\begin{center}
\small(a) \includegraphics[width=7cm]{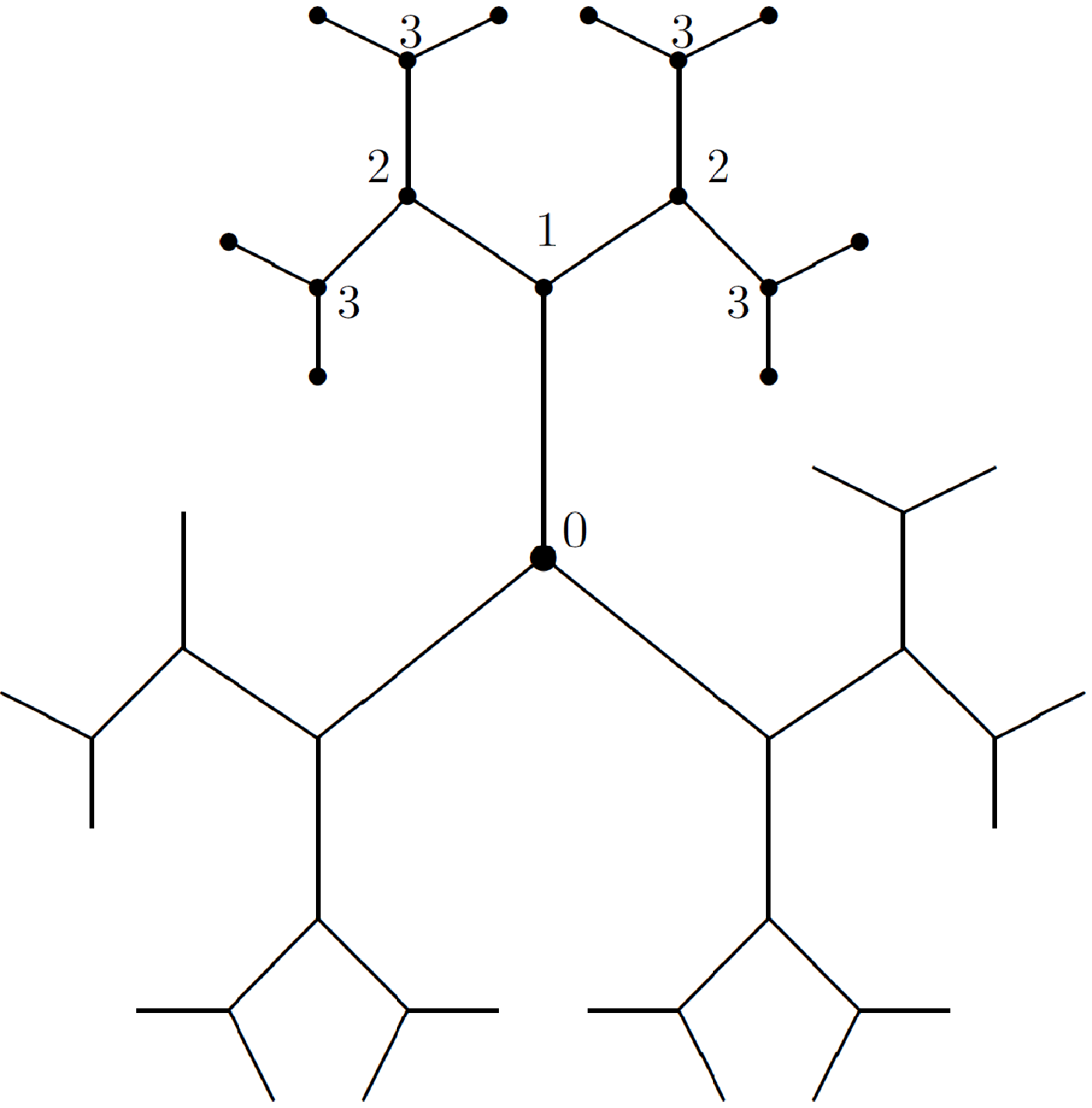}
\small(b) \includegraphics[width=7cm]{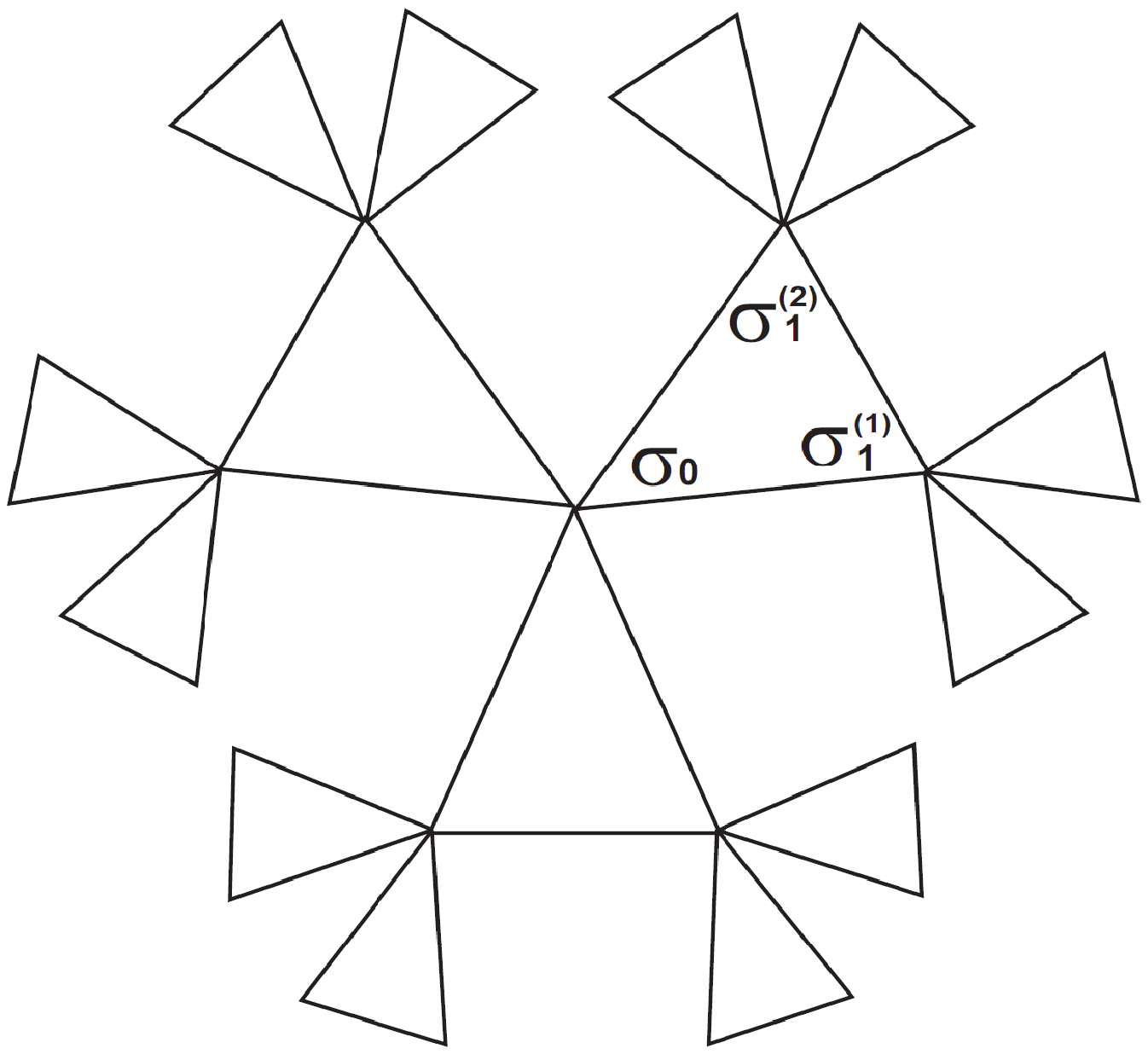}
\caption {\small{The Bethe (a) and the Husimi (b) lattices with coordination number $\gamma=3$. \label{lattice}}}
\end{center}
\end{figure}

The $Q$-state Potts (QSP) model on a Bethe lattice (Fig. \ref{lattice}(a)) and the three-site interaction antiferromagnetic Ising (TSAI) model on a Husimi lattice (Fig. \ref{lattice}(b)) are defined by the following Hamiltonians \cite{pottsham, isingham}:
\begin{eqnarray}\label{3}
{\mathcal{H}_{QSP}}=-J\sum_{(i,j)}\delta({{\sigma_{i},\sigma_{j}} })-H
\sum_{i}\delta({{\sigma_{i},Q} }), \\
{\mathcal{H}_{TSAI}}=-J_3\sum_{\bigtriangleup}\sigma_{i}\sigma_{j}\sigma_{k}-H\sum_{i}\sigma_{i}, \label{4}
\end{eqnarray}
where $\sigma_i=1, 2, ..., Q$ for the Potts model, and $\sigma_i=\pm 1$ for the TSAI model, $\delta(x, y)$ being the Kronecker delta. The first sum in (\ref{3}) and (\ref{4}) runs over the nearest neighboring sites and triangles respectively, while the second one goes over all lattice sites ($J_3<0$ and $J<0$ correspond to the antiferromagnetic case).

The partition function and the single site magnetization of these models are given by
\begin{eqnarray}
\mathcal{Z}_{QSP}=\sum_{\{{\sigma}\}}{e^{-\frac{\mathcal{H}_{QSP}}{k_B T}}},\nonumber
\\ \mathcal{Z}_{TSAI}=\sum_{\{{\sigma}\}}{e^{-\frac{\mathcal{H}_{TSAI}}{k_B T}}};\nonumber
\\ M_{QSP}=\langle\delta(\sigma _0,Q)\rangle=\mathcal{Z}_{QSP}^{-1}\sum_{\{\sigma\}}{\delta(\sigma _0,Q)}e^{-\frac{\mathcal{H}_{QSP}}{k_BT}},\nonumber
\\ M_{TSAI}=\langle\sigma_0\rangle=\mathcal{Z}_{TSAI}^{-1}\sum_{\{\sigma\}}{\sigma _0}e^{-\frac{\mathcal{H}_{TSAI}}{k_BT}},\nonumber
\end{eqnarray}
where we have set $k_B=1$.

By employing the recursion relation technique we separate the Bethe (Husimi) lattice into $\gamma$ identical branches by cutting them apart at the central point (the central triangle). In such a way, by following the well-known procedure \cite{recur, recur1, recur2, jetp, Husimi, Husimi1, He3, He31, gujrati, gujrati1}, we obtain:

\begin{eqnarray}
&M_{QSP}=\langle\delta(\sigma_0, Q)\rangle
=\frac{e^\frac{H}{T}}{e^\frac{H}{T}+(Q-1)x_{n}^\gamma}, \nonumber
\\ &M_{TSAI}=\langle\sigma_0\rangle=\frac{e^{\frac{2 H}{T}} y_n^{\gamma }-1}{e^{\frac{2 H}{T}} y_n^{\gamma }+1}, \nonumber
\end{eqnarray}
where
\begin{eqnarray}
x_n=f_1(x_{n-1}),& \nonumber
\\ f_1(x)=\frac{e^{\frac{H}{T}}+(e^{\frac{J}{T}}+Q-2)x^{\gamma-1}}{e^{\frac{H+J}{T}}+(Q-1)x^{\gamma-1}};& \label{8}
\\ y_{n}=f_2(y_{n-1}),& \nonumber
\\ f_2(y)=\frac{y^{2 (\gamma -1)} e^{\frac{4 H+2 J_3}{T}}+2 e^{\frac{2 H}{T}}
   y^{\gamma -1}+e^{\frac{2 J_3}{T}}}{2 y^{\gamma -1} e^{\frac{2 H+2
   J_3}{T}}+e^{\frac{4 H}{T}} y^{2 (\gamma -1)}+1}.& \label{9}
\end{eqnarray}
Notice that the mapping $f_1(x)$, and thus the QSP model, is well defined for non-integer values of $Q$. A number of physical phenomena, like dilute spin glasses, branched polymers, and self organizing critical systems can be formulated in terms of the model with non-integer $Q$ \cite{lyubensky, whittle, aharony}.

Hereafter we fix the coordination number to $\gamma=3$. The mappings $f_1(x)$ and $f_2(y)$ are maximal at $x=x^*=0$ and $y=y^*=e^{-H/T}$ (the $f_2(y)$ map is considered in the region $y>0$).

\section{Superstable cycles}\label{sup}

In the present section we discuss the superstability properties of the rational mappings (\ref{8}) and (\ref{9}). Generally, the values of parameters, at which a mapping $f(x)$ satisfies the condition
\begin{eqnarray}
\left\{\begin{array}{ll}
f^{(n)}(x)=x & \\
(f^{(n)}(x))'=0, &
\end{array} \right.\label{1}
\end{eqnarray}
form a line in the parameter space (\textit{e.g.} in the ($T;H$) plane), called a superstable $n$-cycle \cite{ecol, lee, kaneko} ($f^{(n)}(x)$ stands for the $n-th$ iteration of a map $f(x)$). Since the Lyapunov exponent is defined as
\begin{eqnarray}
\lambda{(x)}=\lim_{n\rightarrow\infty}{\frac{1}{n}}\ln\left|\frac{df^{(n)}(x)}{dx}\right|, \nonumber
\end{eqnarray}
one finds that at points of superstability $\lambda(x)=-\infty$ (hence the name \textit{superstable}).

The second equation in (\ref{1}) corresponds to the extremum condition of a mapping $f(x)$. Thus, putting the value of $x=x^*$ for $f_1(x)$ and $y=y^*$ for $f_2(y)$ in the first line of (\ref{1}), we obtain an equation for the superstable cycle. The latter can be solved analytically for the QSP model on a Bethe lattice for both $n=2$ and $n=3$, while for $f_2(y)$ this can be done only for $n=2$ (see below).

\subsection{\label{app} Analytic expressions for superstable cycles}
Let $g_1(x)\equiv f_1^{(2)}(x)=f_1[f_1(x)]$. As mentioned above, to obtain a superstable cycle of the $n-th$ order the critical point ($x^*=0$ for $f_1(x)$) is inserted in the first line of (\ref{1}). Thus, we consider the equation $g_1(0)=0$:

\begin{eqnarray}
\frac{e^{h/T}+e^{-\frac{2 J}{T}}
   \left(e^{J/T}+Q-2\right)}{e^{\frac{h+J}{T}}+(Q-1) e^{-\frac{2
   J}{T}}}=0. \label{11}
\end{eqnarray}
The solution of (\ref{11}) is
\begin{eqnarray}
H=T \ln \left(-e^{J/T}-Q+2\right)-2 J: \label{12}
\end{eqnarray}
this yields the superstable cycle of the second order for the QSP model.

By denoting $h_1(x)\equiv f_1^{(3)}(x)$ and by solving the equation $h_1(x^*)=x^*$, one obtains the superstable cycle of the third order. The equation $h_1(0)=0$ is reduced to the following one:

\begin{eqnarray}
z^6 \mu ^3+ a \mu ^2 + b \mu + c=0, \label{13}
\end{eqnarray}
where

\begin{eqnarray}\nonumber
a=z^3 \left(Q z+2 Q+z^2-2 z-2\right),\\ \nonumber
b=Q^2 \left(2 z^2+1\right)+Q \left(4 z^3-8 z^2-2\right)+2 z^4-8 z^3+8
   z^2+1, \\ \nonumber
c=(-2 + Q + z)^3,\\ \nonumber
z=e^{J/T},\\ \nonumber
\mu=e^{H/T}.
\end{eqnarray}
The only real solution of (\ref{13}) corresponds to the superstable cycle of the third order:
 \begin{eqnarray}
\mu=\frac{a^2-a u-3 b z^6+u^2}{3 u z^6},
 \end{eqnarray}
with
\begin{equation}
u=\sqrt[3]{-2 a^3+\sqrt{\left(2 a^3-9 a b z^6+27 c z^{12}\right)^2-4 \left(a^2-3 b z^6\right)^3}+9 a b z^6-27 c z^{12}}. \nonumber
\end{equation}

In a similar way one finds superstable cycles of the TSAI model on a Husimi lattice. In particular, the superstable cycle of the second order can be found from the equation $g_2(e^{-H/T})=e^{-H/T}$ $(g_2(x)\equiv f_2[f_2(x)])$, which is reduced to

\begin{eqnarray}
\frac{k l^2+k+2 l}{2 k l+l^2+1}=\frac{1}{\sqrt{l}},  \label{14}
\end{eqnarray}
where
\begin{eqnarray}\nonumber
k=e^{2J_3/T},\\ \nonumber
l=e^{2H/T}.
\end{eqnarray}
By knowing that the TSAI model always possesses a superstable cycle at a zero magnetic field, we can factorize Eq.~(\ref{14}) with respect to $(l-1)$, obtaining:
\begin{equation}
k^2 l^4+\left(k^2+4 k-1\right) l^3+3 \left(k^2+1\right) l^2+\left(-k^2+4
   k+1\right) l+1=0. \label{15}
\end{equation}
This equation has two real solutions, each corresponding to one of the two superstable cycles of the second order of the TSAI model (superstable orbits of the second order always appear in pairs here). Although being straightforward, the exact solution of (\ref{15}) is too bulky for being presented here.

For higher orders of superstability, equations of the type $f^{(k)}(x^*)=x^*$ (with $k>3$ for the QSP model and $k>2$ for the TSAI model) are solved numerically.

\subsection{Superstability in the period three window}

\begin{figure}[ht]
\begin{center}
\small(a) \includegraphics[width=7cm]{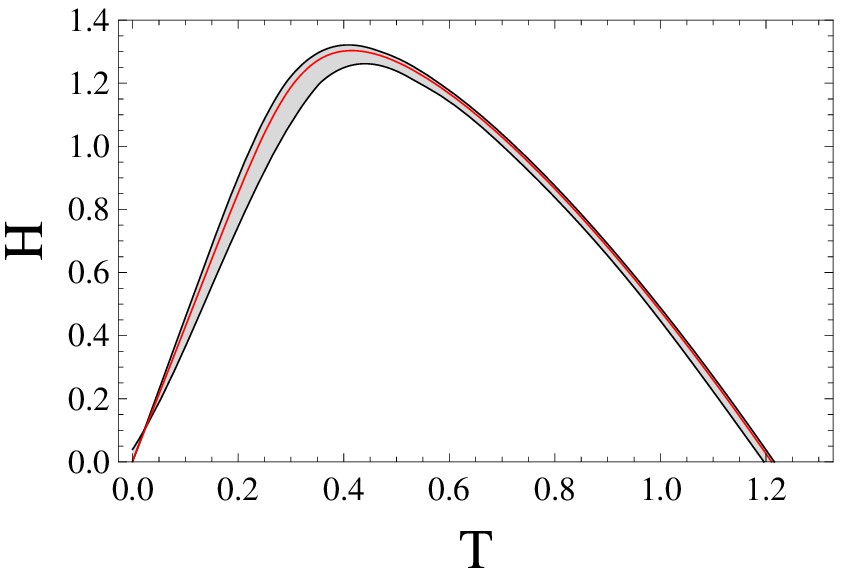}
\small(b) \includegraphics[width=7cm]{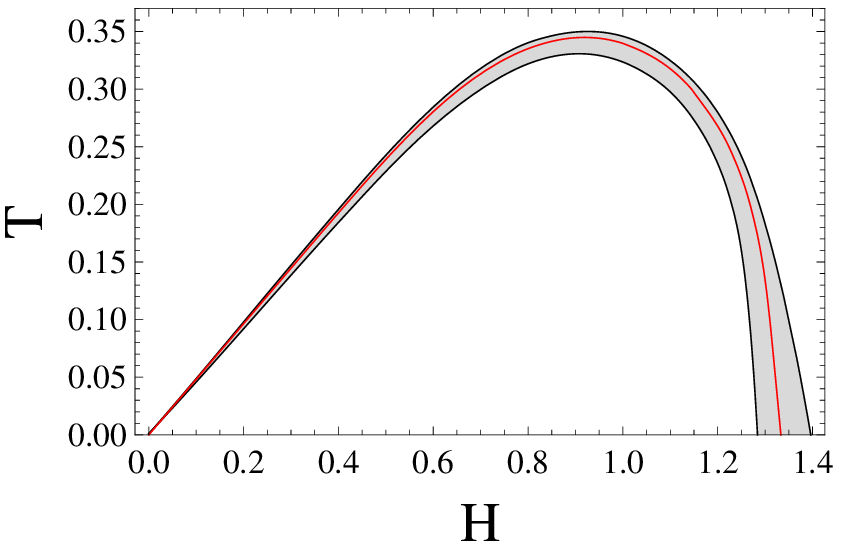}
\caption {\small{The period three modulated phases (shaded regions; results taken from~\cite{jetp}) and superstable cycles of the third order (red -central- curves) for (a) QSP model on a Bethe lattice for $Q=1.1$, $J=-1$, $\gamma=3$ (for positive values of the magnetic field $H$); (b) TSAI model on a Husimi lattice for $J_3=-1$, $\gamma=3$. \label{super}}}
\end{center}
\end{figure}

In the present work we are mainly interested in the period three window of the above described models. In Fig.~\ref{super} we show the period three modulated phase in the $(T; H)$ plane (the shaded region) and the superstable cycle of the third order (the red -central- line).

On the upper black curves in Figs. \ref{super} (a) and (b) a tangent bifurcation  \cite{pom, pom1, pom2} leads to a transition between chaotic and period three modulated phases. The lower black curve corresponds to the transition line between period three and six: here a conventional period doubling occurs (see~\cite{jetp} for details).

\begin{figure}[ht]
\begin{center}
\small(a) \includegraphics[width=7cm]{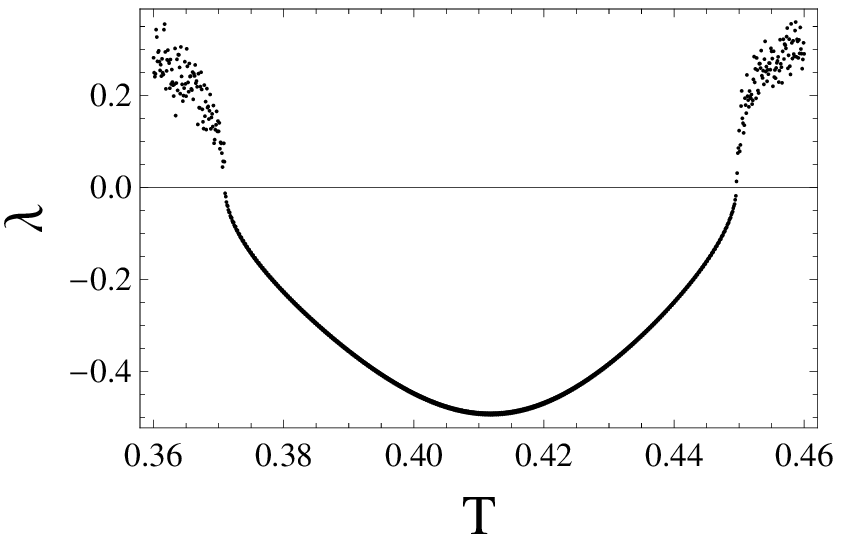}
\small(b) \includegraphics[width=7cm]{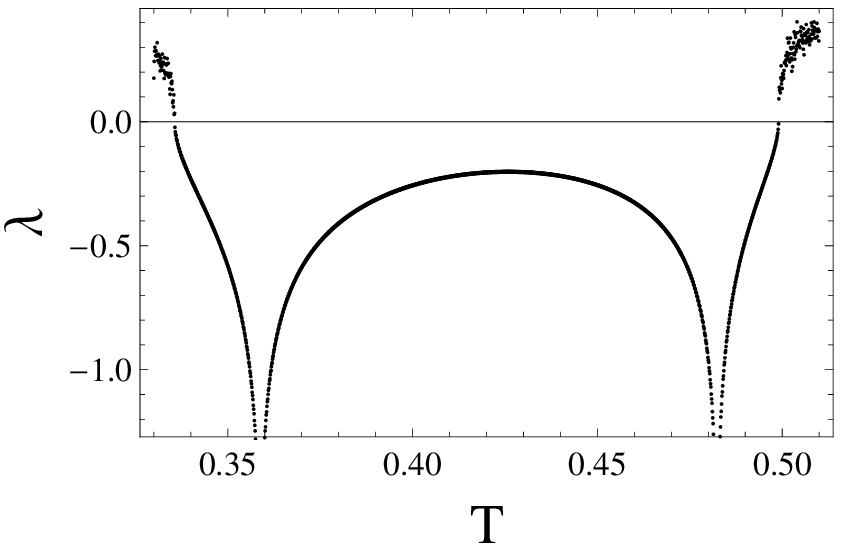}\\
\small(c) \includegraphics[width=7cm]{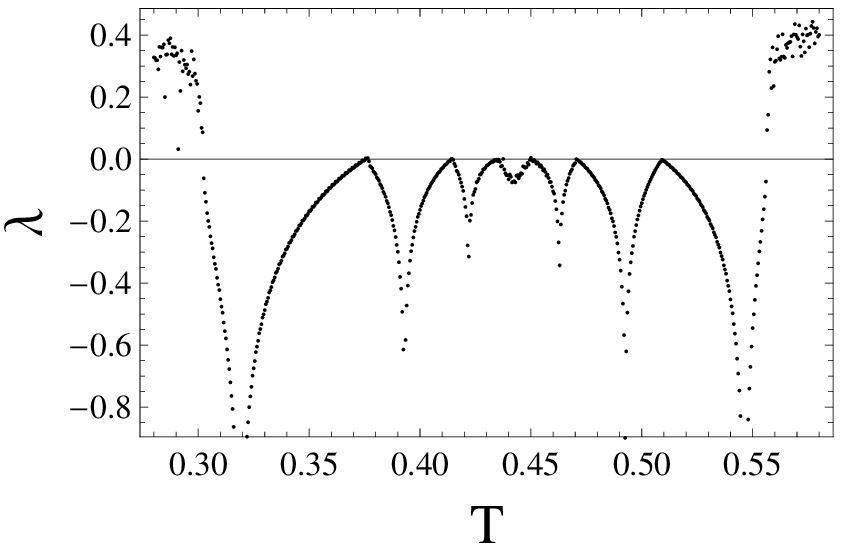}
\small(d) \includegraphics[width=7cm]{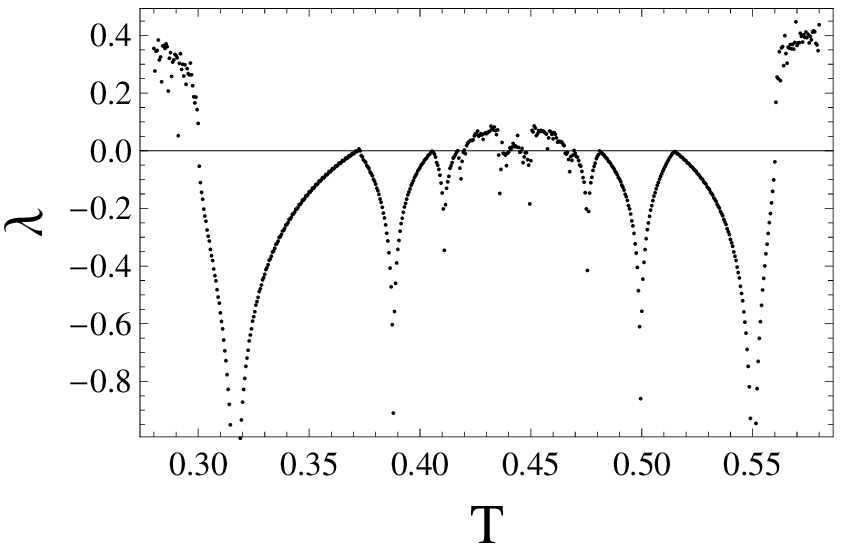}
\caption {\small{ The Lyapunov exponent $\lambda$ in the period three window of the QSP model versus the temperature $T$ for $Q=1.1$, $J=-1$, $\gamma=3$ and (a) $H=1.31$; (b) $H=1.28$;  (c) $H=1.227$; (d) $H=1.223$. \label{lyap}}}
\end{center}
\end{figure}
\begin{figure}[ht]
\begin{center}
\small(a) \includegraphics[width=7cm]{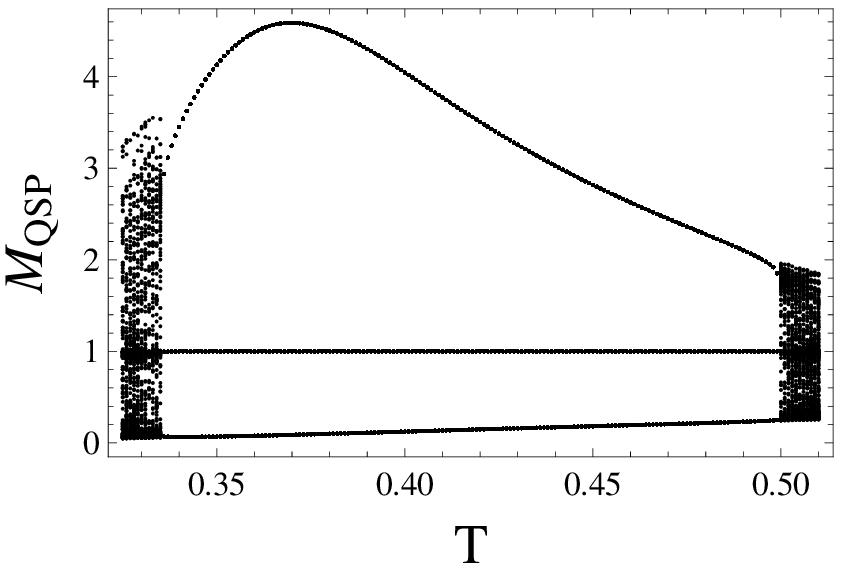}
\small(b) \includegraphics[width=7cm]{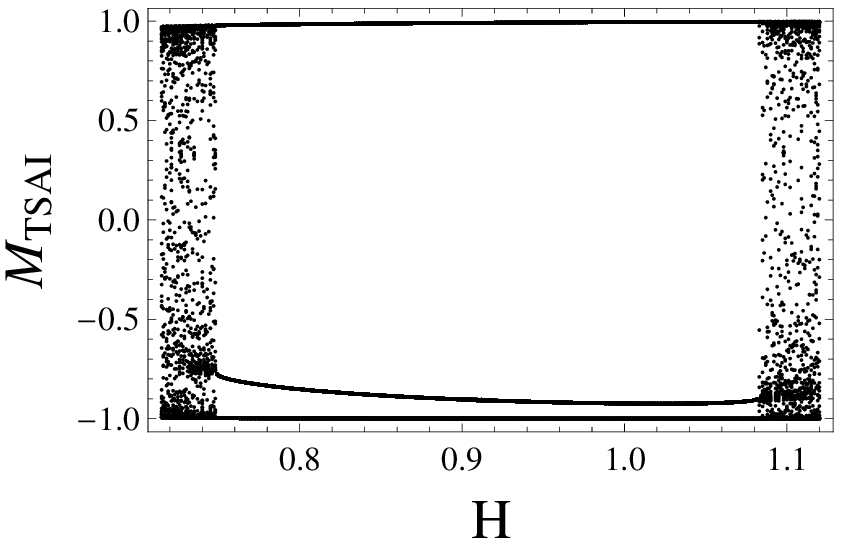}
\caption {\small{The magnetization $M$ in the period three window of (a) QSP model for $Q=1.1$, $J=-1$, $\gamma=3$, $H=1.28$ versus the temperature $T$; (b) TSAI model on a Husimi lattice for $J_3=-1$, $T=0.33$, $\gamma=3$, versus the magnetic field $H$. \label{magn}}}
\end{center}
\end{figure}

Below we consider the region of positive magnetic field for the QSP model on a Bethe lattice ($1<Q<2$), since here it exhibits some interesting features with respect to the temperature $T$: when the line $H=const$ intersects only the chaotic $-$ period three phase transition line (the upper black curve in Fig. \ref{super} (a)), the system does not possess any superstable cycle (Fig. \ref{lyap} (a)). Furthermore, when the line $H=const$ intersects the red curve in Fig. \ref{super} (a), we find two superstable cycles (Fig. \ref{lyap} (b)). {We emphasize that these superstable cycles are not confined between any doubling bifurcations. Only after the $H=const$ line intersects the lower black curve of Fig. \ref{super} (a), more \textit{pairs} of superstable cycles, confined between doubling bifurcations appear (Fig. \ref{lyap} (c)).} Finally, at weaker magnetic field we observe a chaotic regime, confined inside the period three window (Fig. \ref{lyap} (d)). The TSAI model on a Husimi lattice behaves similarly with respect to the magnetic field $H$ (at a fixed temperature $T$). Note that we find neither quantitative, nor qualitative changes of the thermodynamic properties at points of superstability (compare e.g. Figs. \ref{magn} (a) and \ref{lyap} (b)). At points of superstability we find a change in the symbolic dynamics, as we will discuss in what follows.

We emphasize a non-trivial connection of bifurcation and superstability in the periodic window: in some parameter regions a superstable cycle is not followed by a doubling bifurcation ($T\in[0.331, 0.35]$ for the Ising model, and $H\in [1.261, 1.321]$ for the Potts model with $Q=1.1$). This behavior is different from that of the logistic map, where, by increasing the nonlinearity parameter, a superstable cycle is always followed by a doubling bifurcation. We ascribe this distinction to the following fact: in our case (QSP model with respect to $T$ for $H>0$, and TSAI model with respect to $H$), the physical parameters vary in such a way that period three windows have strictly distinguishable edges, at which a tangent bifurcation takes place \cite{pom, pom1}. This is indeed quite different from simply increasing the nonlinear parameter in the logistic map: the tangent bifurcation there occurs only at one edge of the window, with a full period doubling cascade afterwards. Somehow similar results were recently pointed out in~\cite{diam, kenna}, where a superstable cycle without emergence of any doubling bifurcation was found. Nevertheless, in the conventional, $1\cdot 2^n$ period doubling cascade, the models exhibit the usual relation between superstability and bifurcation, as that of the logistic map (the same for the Potts model with respect to the magnetic field at a fixed $T$, and for the Ising model with respect to the temperature at a fixed $H$, {\it in the periodic window}).

In Fig. \ref{super} (b) one finds a region of the magnetic field $H\in[1.276; 1.375]$, where the ground state of the system is the period three modulated phase. Additionally, in the region $H\in[1.338; 1.375]$ the TSAI model does not exhibit any superstable cycle of the third order with respect to the temperature $T$ (when $H=const$).

\section{\label{symb}Symbolic dynamics and superstability}

In this section we discuss the qualitative changes in the properties of mappings (\ref{8}) and (\ref{9}), that occur at points of superstability. In order to accomplish that we perform a detailed analysis of the associated symbolic dynamics \cite{dav}, by using the conventional binary partition for unimodal maps: $x^*$ denotes the critical point, $R$ the region to its right and $L$ the leftmost region.  We consider values of $Q$ in the interval $1<Q<2$, to have a chaotic regime and to avoid the divergence of $f_1(x)$.

\begin{figure}[!h]
\begin{center}
\small{(a)} \includegraphics[width=15cm]{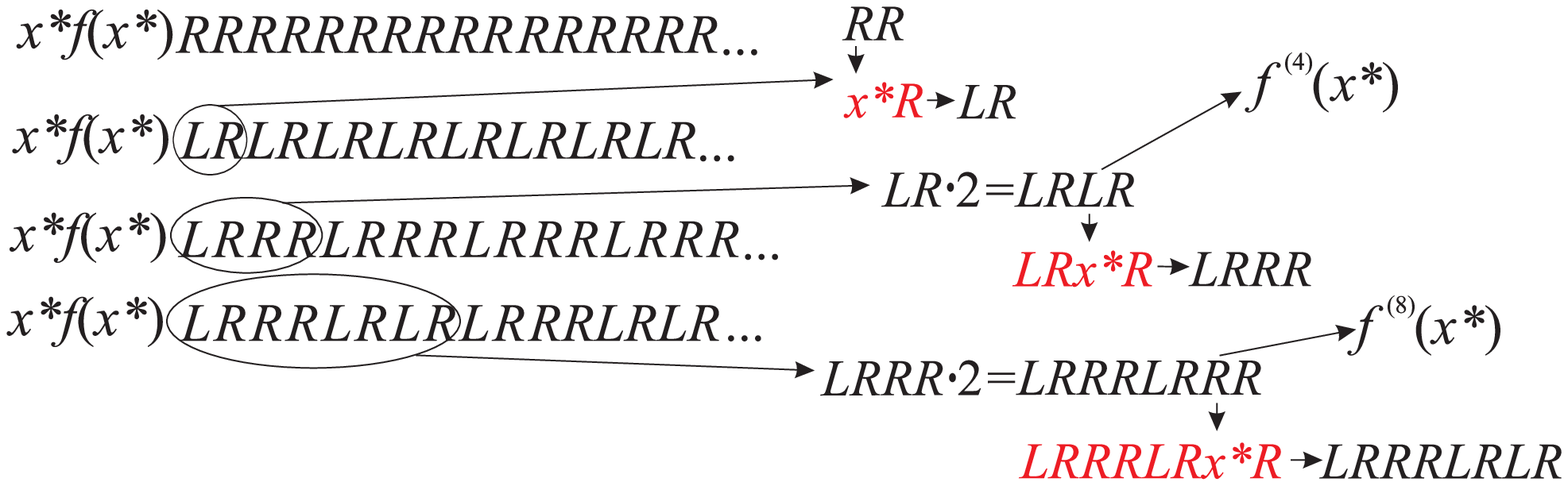} \\
\small{(b)} \includegraphics[width=15cm]{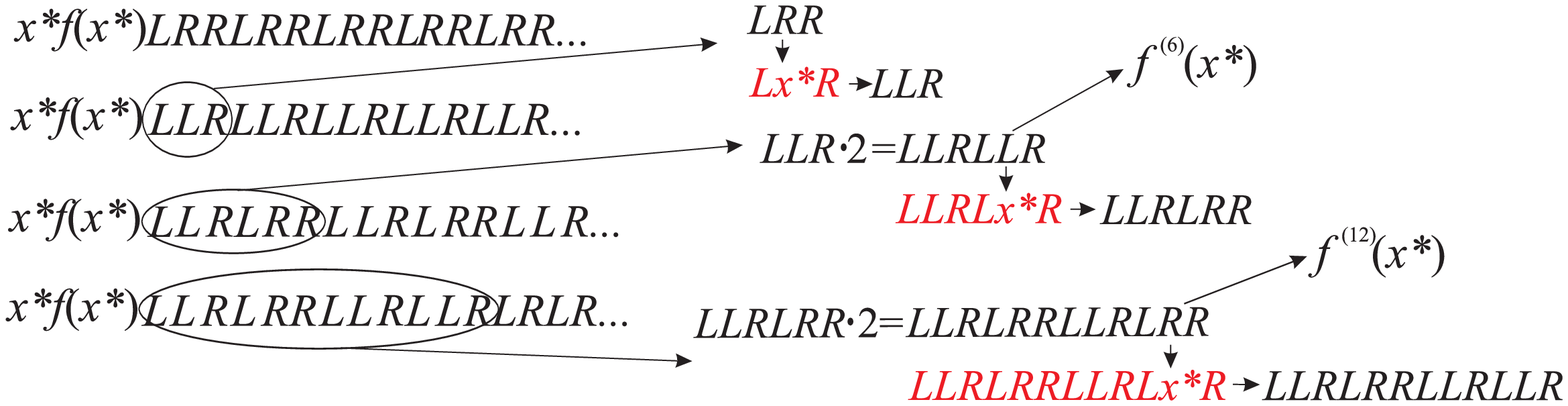}
\caption {(a) The generic $(R; L)$ sequence for the mappings of the QSP and TSAI models (both denoted by $f(x)$ for simplicity) in the \textit{conventional period doubling} regime. $x^*$ is the extremum point: for the QSP mapping we have $x^*=0$, while for the TSAI mapping $x^*=e^{-\frac{H}{T}}$. Red (grey) color defines the $(R; L)$ sequence \textit{at} a superstable point, while the circles identify the repetitive block. Each new sequence appears after a superstable point of a higher order than the previous one. The penultimate character of a doubled block is the $n-th$ iteration of the mapping $f(x)$. (b) The same, but for the \textit{period three window}. \label{RL2}}
\end{center}
\end{figure}

Figure \ref{RL2} shows the changes, that the symbolic $(R; L)$ sequence of mappings $f_1(x)$ and $f_2(x)$ undergo at points of superstability. We find that in order to obtain each next $(R; L)$ sequence, one has to take the repetitive block of the previous one, doubles it and replaces the penultimate $L$ with $R$ (or vice-versa). Therefore, a ($R; L$) sequence can be constructed for the full doubling cascade in a recursive way, by simply identifying the first repetitive block. In particular, for the conventional period doubling regime, this block is simply $RR$ (of length $2$), while for the 3 window it is $LRR$ (of length 3). Generally, for any period-$p$ window, the first block is of a length $p$. Note that in the sequence for a superstable point, the above mentioned penultimate character has to be replaced by $x^*$ (red -grey- sequence in Fig. \ref{RL2}).

The reason why the {\it penultimate} character changes at points of superstability becomes clear from the following discussion: as known, at a superstable cycle of the $n-th$ order the extremum point $x^*$ of a mapping belongs to the cycle, so that $f^{(n)}(x^*)=x^*$. Meanwhile, the $n-th$ iteration of a mapping corresponds to the penultimate character of a doubled block (see Fig. \ref{RL2}): so that symbol $x^*$ at a point of superstability.

It is worth to mention, that the generic $(R; L)$ sequence can distinguish the regions between two consecutive superstable cycles. Thus, we can introduce a notion of a ``phase" with a fixed symbolic sequence. For instance, in the case of Fig. \ref{lyap} (a), the period three window features only one $LRR$ ``phase". As for Fig. \ref{lyap} (b), here we have two phases and two transitions between them: $LRR\rightarrow LRRLLR$ at the first superstable point (of the third order) and a backwards transition $LRRLLR\rightarrow LRR$.
We do not find here a transition to the next sequence, as in Fig.~\ref{RL2}(b),
since both superstable points above are of the same, third order
%
%In fact, the areas before the first superstable point and after the second one are the same ``phases", since they are described by
%identical symbolic sequences.
(symbolic dynamics evolves by following the rules of Fig.~\ref{RL2} if we have a transition through a superstable orbit of a higher order). Finally, for lower values of the magnetic field we have more phases of a higher periodicity and two-way transitions between them. Meanwhile, in the conventional period doubling regime the $(R; L)$ sequence forms a devil's staircase with a monotonic behavior (the same for the period three window with respect to the magnetic field (temperature) for the Potts (Ising) model).

We additionally note, that the present method of tracking the symbolic dynamics at superstable points remains true for the logistic map as well.

\section{\label{concl}Discussion and conclusions}

In this work we studied the superstability phenomenon in two examples of rational mappings, describing real statistical systems: the $Q$-state Potts model on a Bethe lattice and the three-site antiferromagnetic Ising model on a Husimi lattice. We provided an exact solution for the superstable cycles of the second order for both models. Additionally, we presented an analytical result for the superstable orbit of the third order (in the period three window) for the rational mapping describing the Potts model (for the Ising model we had to rely on numerical analysis). Notice that the list of exact results for period three windows is quite short and it includes solutions mainly for the logistic (i.e. polynomial) map. The bulk of the paper was concentrated on the period three window, since here both models feature some interesting properties. Namely, we found that in some regions of the temperature $T$ and the magnetic field $H$ a superstable cycle is not confined between doubling bifurcations. This is the consequence of an intrinsic phase structure of both systems, and particularly of the fact that the period-3 window has two distinguishable edges. A tangent bifurcation occurs at both of them, so that by varying a physical parameter the scenario is quite different from the logistic map bifurcation structure.

On the other hand, we analyzed the changes in mapping properties at points of superstability, by using symbolic dynamics technique. We pointed out how the generic ($R; L$) sequence is affected at points of superstability. Furthermore, we showed that the ($R; L$) sequence after each consecutive superstable orbit can be deduced from the previous one in a straightforward way. We also observed that this technique works in any periodic regime of the above described rational mappings and remains true for polynomial maps.

\section*{Acknowledgments}

We thank Professor Stefano Ruffo for helpful discussions and useful comments. This work was supported by Marie Curie ``DIONICOS" program under the proposal of FP7$th$ PIRSES-GA-2013-612707 and SCS MES RA  in the frame of the research project No. SCS 13-1C137 grants. L.C. gratefully acknowledges funding by the Regional Council of Burgundy (Conseil R\'{e}gional de Bourgogne).

\end{document}